\documentclass[twocolumn,showpacs,preprintnumbers,amsmath,amssymb]{revtex4}

\usepackage{graphicx}
\usepackage{dcolumn}
\usepackage{bm}

\begin{document}

\title{Entanglement of scattered single photon with atom}

\author{Rui Guo}
\author{Hong Guo}\thanks{Author to whom correspondence should be
addressed. E-mail: hongguo@pku.edu.cn, phone: +86-10-6275-7035, Fax:
+86-10-6275-3208.}

\affiliation{CREAM Group, Key Laboratory for Quantum Information and
Measurements of Ministry of Education, School of Electronics
Engineering $\&$ Computer Science, Peking University, Beijing
100871, P. R. China\\}%

\date{\today}

\begin{abstract}
Single--photon which is initially uncorrelated with atom, will
evolve to be entangled with the atom on their continuous kinetic
variables in the process of resonant scattering. We find the
relations between the entanglement and their physical control
parameters, which indicates that high entanglement can be reached by
broadening the scale of the atomic wave or squeezing the linewidth
of the incident single--photon pulse.
\end{abstract}

\pacs{03.65.Ud, 42.50.Vk, 32.80.Lg }.

\maketitle

\section{Introduction}
Quantum entanglement is of fundamental importance in the theory of
quantum nonlocality\cite{1 nonlocality} as well as in quantum
information\cite{2 QIT}. Recently, photon--atom entanglement is
frequently discussed in their finite Hilbert spaces\cite{3 etagl
fini}, such as, the polarizations of photon or the internal states
of atom. With the progress of micro--cavity quantum
electrodynamics\cite{4 CQED} and high coupling artificial
atom\cite{5 artificial atom}, single photon raises its ability to
affect considerably not only the atom's internal state but also its
external motion. As a result, it gives rise to some basic questions
related to the photon--atom entanglement on their infinite kinetic
degree of
freedom.\\

In recent studies\cite{6 Singlephoton}\cite{7 scattering},
entanglement in the continuous kinetic variables between
single--photon and atom is mostly discussed in the process of
single--photon emission with atomic recoil, where the atom is
initially pumped to its excited level and the single--photon is
prepared ``intrinsically'' by the atomic spontaneous emission. In
our work, however, the resonant single--photon is initially injected
from a tuneable single--photon generator\cite{8 singlephoton
generator}, whereas an artificial atom is placed freely in vacuum on
its steady state (``artificial'' indicates that the atomic coupling
to the single--photon is stronger than usual, which ensures the
interaction observable\cite{statement}). We find that, after the
interaction, the scattered single--photon will be entangled to the
atom at a higher degree compared with the case of solely spontaneous
emission. We explain this phenomena as the coherent pumping of the
incident photon and evaluate it with a defined ``entanglement
pumping coefficient''.\\

To describe the degree of entanglement, firstly, we use the ratio
($R$) between the conditional and unconditional variance in momentum
to evaluate the two particles' correlation in the probability
amplitude of their wave function, which is experimentally accessible
and can be seen as the ``amplitude entanglement'' in momentum
space\cite{9 photoionization}\cite{10 phase ent}; secondly, we use
the standard Schmidt decomposition\cite{11 Schmidt dec} and treat
Schmidt number $K$\cite{12 Schmidt num} as a criterion for the full
entanglement contained both in amplitude and phase. For both
criterions $R$ and $K$, we revealed their dependencies on the
physical control parameters $\tau$ and $\eta$, and compare them in
some region of interests, from which it is shown that: higher
entanglement can be achieved by either broadening the scale of the
atomic wave or squeezing the linewidth of the incident
single--photon. Transmitted photon is also considered, which is
different to the scattered photon, and exhibits little entanglement
with the atom due to its interference with the transparent wave (initially incident photon wave profile).\\

\begin{figure}
\centering
\includegraphics[height=6cm]{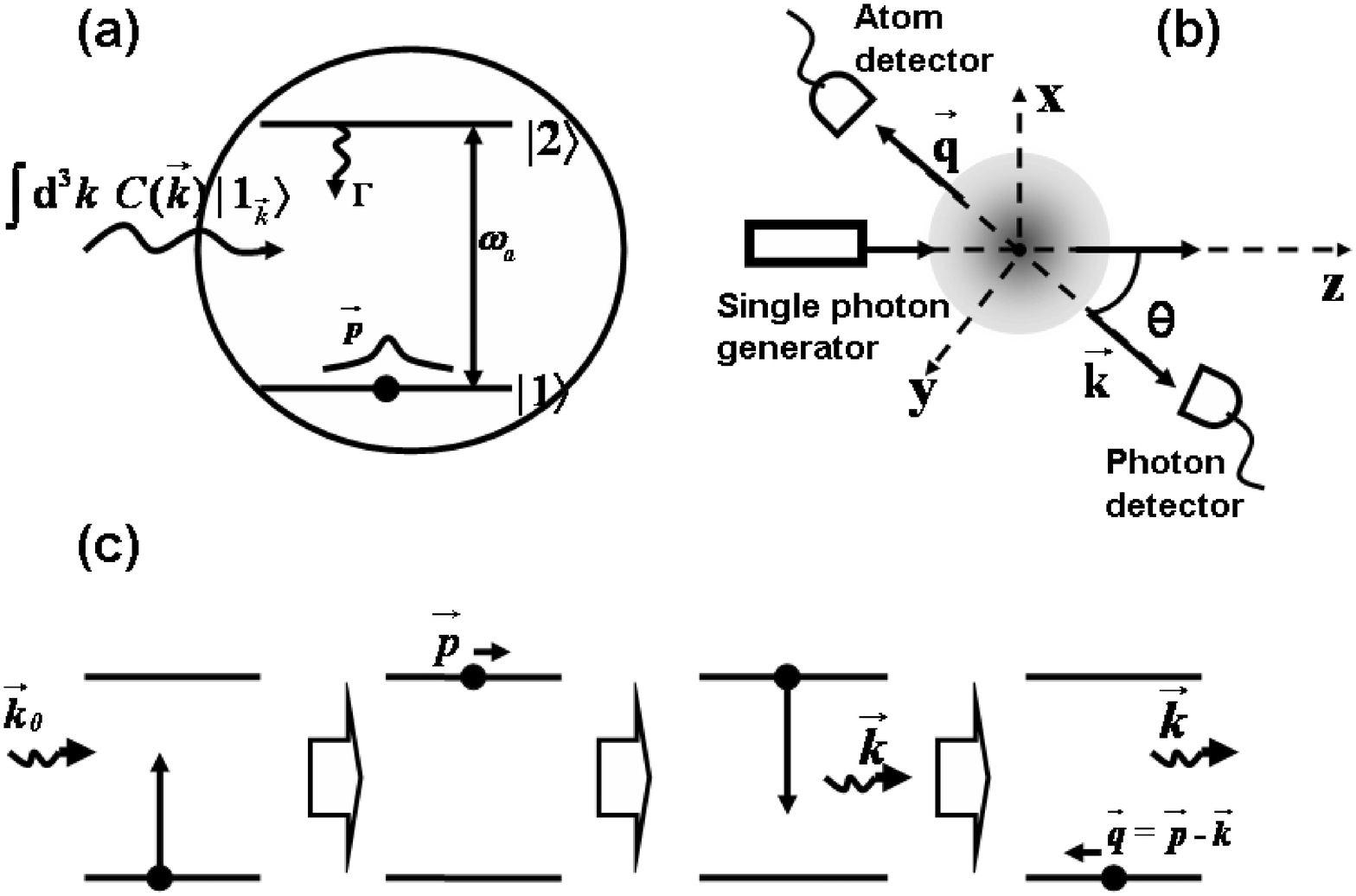}
\caption{(a) Single--photon interacts resonantly with free
two--level atom.\\
(b) The incident photon is scattered by the atom, angle $\theta$ is
fixed to determine the
direction of the detection.\\
(c) Schematic diagram for the absorption--emission process. The
process of emission with atomic recoil will generate entanglement
between the recoiled atom and the scattered photon due to the
momentum conservation.}
\end{figure}

\section{theoretical analysis}

As shown in Fig. 1 (a), the two--level atom with transition
frequency $\omega_{a}$ and mass
 $m$ is placed freely in vacuum, the ground and excited states of which are denoted
by $|1\rangle$ and $|2\rangle$, respectively. The incident
single--photon from some generator is resonant with the atom and
exhibits a superposed state of different fock states due to its
linewidth. For the realistic consideration in some
experiments\cite{13 exp1}, we fix the photon detector and atom
detector in opposite directions and make them both in the $x$--$z$
plane for simplicity as in Fig. 1 (b), the angle $\theta$ can be
chosen to observe the scattering in needed directions. \\

Under the rotating wave
  approximation (RWA) the Hamiltonian can
 be written in Schr\"{o}dinger picture as:
\begin{eqnarray}
\nonumber\hat{H}&=&\frac{(\hbar
\hat{p})^{2}}{2m}+\sum_{\vec{k}}\hbar
\omega_{\vec{k}}\hat{a}^{\dag}_{\vec{k}}\hat{a}_{\vec{k}}+\hbar
\omega_{a}\hat{\sigma}_{22}\\&+&\hbar\sum_{\vec{k}}\left[g(\vec{k})\hat{\sigma}_{12}\hat{a}^{\dag}_{\vec{k}}e^{-i\vec{k}\cdot\vec{r}}
+{\rm H.c.} \right],
\end{eqnarray}
where $\hbar \hat{p}$ and $\vec{r}$ denote atomic center--of--mass
momentum and position operators, $\hat{\sigma}_{ij}$ denotes the
atomic operator $|i\rangle \langle j|$ ($i,j=1,2$),
$\hat{a}_{\vec{k}}$ and $\hat{a}^{\dag}_{\vec{k}}$ are the
annihilation and creation operators for the light mode with photonic
wave vector $\vec{k}$ and frequency $\omega_{\vec{k}}=ck$,
respectively. Note the summation is performed over all coupled modes
in the continuous Hilbert space. We also suppress the polarization
index in the summation as well as in photon state, since we can
always choose a particular
polarization to detect the photon.  $g(\vec{k})$ is the dipole coupling coefficient.\\

As there is only one photon in the interaction, the basis of the
Hilbert space can be denoted as $|\vec{q},1_{\vec{k}},i\rangle\ \
(i=1,2)$, where the arguments in the kets denote, respectively, the
wave vector of the atom, and of the photon, and the atomic internal
state. At time $t$ the state vector can therefore be expanded as:
\begin{eqnarray}
|\psi\rangle=\sum_{\vec{q},\vec{k}}C_{1}(\vec{q},\vec{k},t)|\vec{q},1_{\vec{k}},1\rangle+\sum_{\vec{q}}C_{2}(\vec{q},t)
|\vec{q},0,2\rangle .
\end{eqnarray}

Substituting Eqs. (1) and (2) into Schr\"{o}dinger equation yields:
\begin{eqnarray}
&&i \dot{A}(\vec{q},\vec{k},t)= g(\vec{k})B(\vec{q}+\vec{k},t)e^{i[ck-\omega_{a}-\frac{\hbar}{2m}(2\vec{q}+\vec{k})\cdot
\vec{k}]t},\\
\nonumber &&i
\dot{B}(\vec{q},t)=\sum_{k}g^{*}(\vec{k})A(\vec{q}-\vec{k},\vec{k},t)e^{i[\omega_{a}-ck+\frac{\hbar
}{2m}(2\vec{q}-\vec{k})\cdot\vec{k}]t},\\
\
\end{eqnarray}
where $A$, $B$ are the slowly varying parts of $C_{1}$ and $C_{2}$,
i.e.:
\begin{eqnarray}
A(\vec{q},\vec{k},t)&=&C_{1}(\vec{q},\vec{k},t)e^{i(\frac{\hbar
q^{2}}{2m}+ck)t},\\
B(\vec{q},t)&=&C_{2}(\vec{q},t)e^{i(\frac{\hbar
q^{2}}{2m}+\omega_{a})t}.
\end{eqnarray}

Suppose the atom is initially in the ground state and has zero
average velocity, the initial condition can be set as:
\begin{eqnarray}
&&A(\vec{q},\vec{k},t=0)=\chi_{0}G(\vec{q})P(\vec{k}-\vec{k_{0}}),\\
&&B(\vec{q},t=0)=0,
\end{eqnarray}
where $G(\vec{q})=G_{x}(q_{x})G_{y}(q_{y})G_{z}(q_{z})$ and
$P(\vec{k}-\vec{k}_{0})=P_{x}(k_{x})P_{y}(k_{y})P_{z}(k_{z}-k_{0})$.
In this case, functions $G_{i}(q_{i})$ and $P_{i}(k_{i})$
$(i=x,y,z)$ have zero center value and bandwidths $\delta q_{i}$ and
$\delta k_{i}$ separately. The coordinates are chosen as in Fig. 1
(b), where we make the incident direction as $z$--axis. $\chi_{0}$
is the normalized factor and
$\vec{k}_{0}=(0,0,\frac{\omega_{a}}{c})$ is
the resonant wave vector.\\

We proceed to solve the equations with Laplace transformation and
single pole approximation\cite{14 single pole} and yield:
\begin{eqnarray}
&&B(\vec{q},t)=-i\chi_{0}\sum_{\vec{k}}g^{*}(\vec{k})\times\\
\nonumber &&\frac{G(\vec{q}-\vec{k})P(\vec{k}-\vec{k}_{0})\left \{
e^{i[\omega_{a}-ck+\frac{\hbar}{2m}(2\vec{q}-\vec{k})\cdot
\vec{k}]t}-e^{-iLt-\Gamma t} \right\}}{iL+\Gamma+i[
\omega_{a}-ck+\frac{\hbar}{2m}(2\vec{q}-\vec{k})\cdot\vec{k}]},
\end{eqnarray}
where the frequency shift $L$ and atomic linewidth $\Gamma$ are
given as:
\begin{eqnarray*}
L&=&\sum_{\vec{k}}\frac{|g(\vec{k})|^{2}}{\omega_{a}-ck+\frac{\hbar}{2m}(2\vec{q}-\vec{k})\cdot
\vec{k}}\ ,
\\ \Gamma&=&\pi
\sum_{\vec{k}}|g(\vec{k})|^{2}\delta(\omega_{a}-ck).
\end{eqnarray*}
We can simplify Eq. (9), by replacing the term $G(\vec{q}-\vec{k})$
with $G(\vec{q}-\vec{k}_{0})$ since the momentum bandwidth $\delta
q_{i}$ due to the recoil is normally much larger than the photon
linewidth $\delta k_{i}$; also, we can replace $\vec{k}$ with
$\vec{k}_{0}$ in the term $\frac{\hbar
}{2m}(2\vec{q}-\vec{k})\cdot\vec{k}$. With these approximations, the
first term in the curly bracket can be seen as the antifourier
transform of the product of the photonic shape and Lorentzian shape,
and will cause a decay at a time scale ${\rm
max}\{\frac{1}{\Gamma},\frac{1}{c\delta k_i}  \}$; the second decay
term $e^{-iLt-\Gamma t}$ is due to the spontaneous emission. Then,
one can directly find that $B(\vec{q},t\rightarrow
\infty)\rightarrow 0$. In the further calculations, we ignore the
frequency shift since it can be treated as a modification of the
atomic transition frequency, and
regard the slowly varying function $g(\vec{k})$ as a constant.\\

With the approximations mentioned above, from Eqs. (3) and (9), we
obtain the steady solution of $A(\vec{q},\vec{k},t\rightarrow
\infty)$:
\begin{widetext}
\begin{eqnarray}
\nonumber A(\vec{q},\vec{k},t\rightarrow
\infty)=&&\chi_{0}G(\vec{q})P(\vec{k}-\vec{k}_{0})
+\frac{\chi_{0}|g|^{2}
G(\vec{q}+\vec{k}-\vec{k}_{0})}{\Gamma-i\left[
ck-\omega_{a}-[\hbar(\vec{q}+\vec{k})]^{2}/2m\hbar+(\hbar
\vec{q})^{2}/2m\hbar \right]}\times \\
&&\sum_{\vec{k}_{1}}\frac{P(\vec{k}_{1}-\vec{k}_{0})}{ i\left[
ck-ck_{1}-\frac{\hbar}{2m}(2\vec{q}+\vec{k})\cdot\vec{k}+\frac{\hbar}{m}(\vec{q}+\vec{k})\cdot\vec{k}_{0}
-\frac{\hbar}{2m}k^{2}_{0}  \right]}\ \ .
\end{eqnarray}
\end{widetext}
From Eq. (10), one sees that the final state is a superposition of
the transparent wave (initially incident photon wave profile,
depicted by the first term on the r.h.s.) and scattering wave
(second term on the r.h.s.). In the scattering part, the atom and
the photon are entangled due to the process of photon absorption and
emission with atomic recoil, which is sketched in Fig. 1 (c). One
may find that the Lorentzian--Gaussian factor in the scattering part
is very similar to that in  the case of spontaneous emission with
recoil\cite{6 Singlephoton}, where the Gaussian term is a reflection
of momentum conservation and
the Lorentzian term indicates the energy conservation.\\

The general formula (10) can be used to analyze the photon scattered
in different directions. Without loss of physical generality, we
choose the initial conditions for the atom as
$G_{i}(q_{i})=e^{-(q_{i}/\delta q_{i})^{2}}$, and for the photon
$P_{i}(k_{i})=1/\left(k_{i}/\delta k_{i}+1\right)$ which is exactly
the case if the incident single--photon is generated by spontaneous
emission. As a remark, we point out that all the conclusions in the
following keep available when the incident photon is chosen to be
other shapes such as Gaussian or whatever.

\section{Amplitude Entanglement in Scattered photon}

\begin{figure}
\centering
\includegraphics[height=8cm]{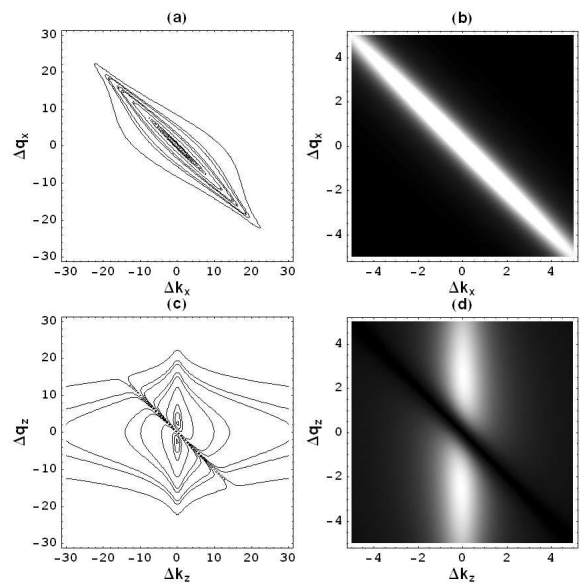}
\caption{(a) and (b) are contour and density plots of
$|A_{\frac{\pi}{2}}|^{2}$ with the condition $\tau_{z}=1$,
$\eta_{x}=10$;
         (c) and (d) are contour and density plots of $|A_{0}|^{2}$ with the condition $\tau_{z}=1$, $\eta_{z}=10$. }
\end{figure}
To make the physical results more evident and avoid unnecessary
mathematical complexity, we focus our attention on the photon
scattered perpendicular to the incident direction, i.e.,
$\theta=\frac{\pi}{2}$. Then we project Eq. (10) into the subspace
$|(q_{x},0,0)\rangle \otimes |1_{(k_{x},0,0)}\rangle$, with the same
approximations used in Eq. (9), and yield:
\begin{eqnarray}
\nonumber A_{\frac{\pi}{2}}&=&\frac{N\cdot {\rm exp}\left[-(\Delta
q_{x}-\frac{\hbar k_{0}}{mc}\Delta
k_{x})^{2}/\eta_{x}^{2}\right]}{(\Delta k_{x}+\Delta
q_{x}+\frac{\hbar k_{0}^{2}}{2m\Gamma}+i)\left[(\Delta k_{x}+\Delta
q_{x})/\tau_{z}+i\right]},\\
&\approx&\frac{N\cdot {\rm exp}\left[-(\Delta
q_{x}/\eta_{x})^{2}\right]}{(\Delta k_{x}+\Delta
q_{x}+i)\left[(\Delta k_{x}+\Delta q_{x})/\tau_{z}+i \right]},
\end{eqnarray}
where $\Delta k_{i}\equiv \frac{k_{i}-k_{0}}{\Gamma/c}$, $\Delta
q_{i}\equiv \frac{\hbar k_{0}}{m\Gamma}(q_{i}-k_{0})$,
$\eta_{i}\equiv \frac{\delta q_{i}\hbar k_{0}}{m\Gamma}$,
$\tau_{i}\equiv \frac{\delta k_{i}}{\Gamma/c}$ $(i=x,y,z)$ are all
defined dimensionless parameters. Note that $\eta_{x}$ and
$\tau_{z}$ contain all the physical parameters that determine the
nature of the atom--photon system, thus can be treated as physical
control parameters for the atom and the photon, respectively. We
neglect tiny terms in Eq. (11) due to $\hbar k^{2}_{0}\ll m\Gamma$
and $\hbar k_{0}\ll mc$ in realistic conditions. $N$ is the
normalization factor where
$N^{2}=\sqrt{2}(1+\tau_{z})/\pi^{\frac{3}{2}}\tau_{z}\eta_{x}$.\\

From Eq. (11) and Fig. 2, one sees that, variables $\Delta q_{x}$
and $\Delta k_{x}$ play the symmetric role in the two Lorentzian
functions. It makes the probability amplitude
$|A_{\frac{\pi}{2}}|^{2}$ localized along the diagonal of the
momentum space, which implies the nonfactorization of the
photon--atom wave function, and then will generate entanglement
between the two particles. In fact, we can treat the ratio ($R$) of
the conditional and unconditional variances for $\Delta q_{x}$ or
$\Delta k_{x}$ as an evaluation of entanglement\cite{9
photoionization}. This ratio, compared to the Schmidt number $K$,
reveals more obvious analytic dependence for the entanglement on its
control parameters $\eta_{x}$ and $\tau_{z}$ , and is also
experimentally directly
accessible\cite{15 exp3}.\\
\begin{figure}
\includegraphics[height=3.5cm]{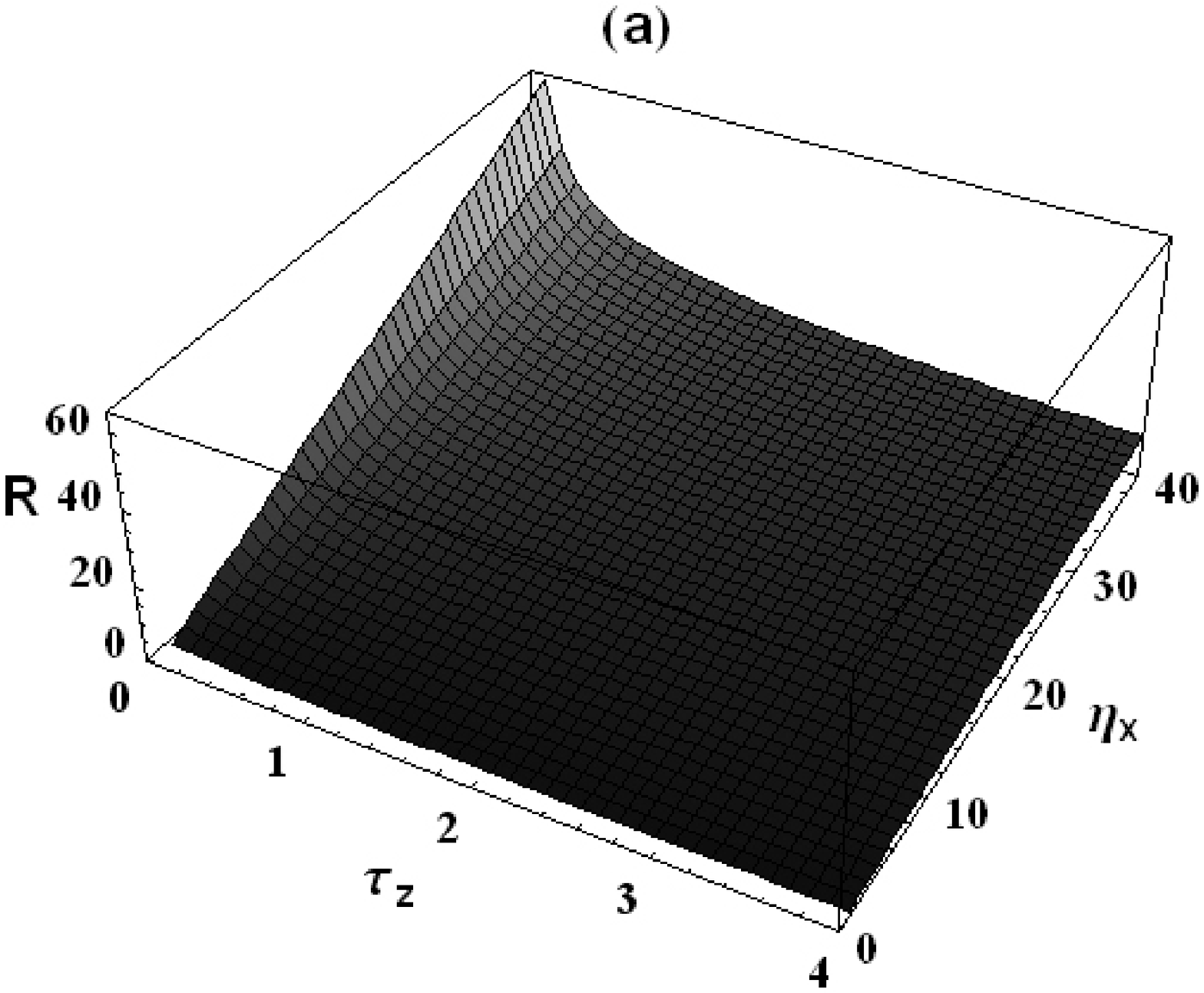}\includegraphics[height=3cm]{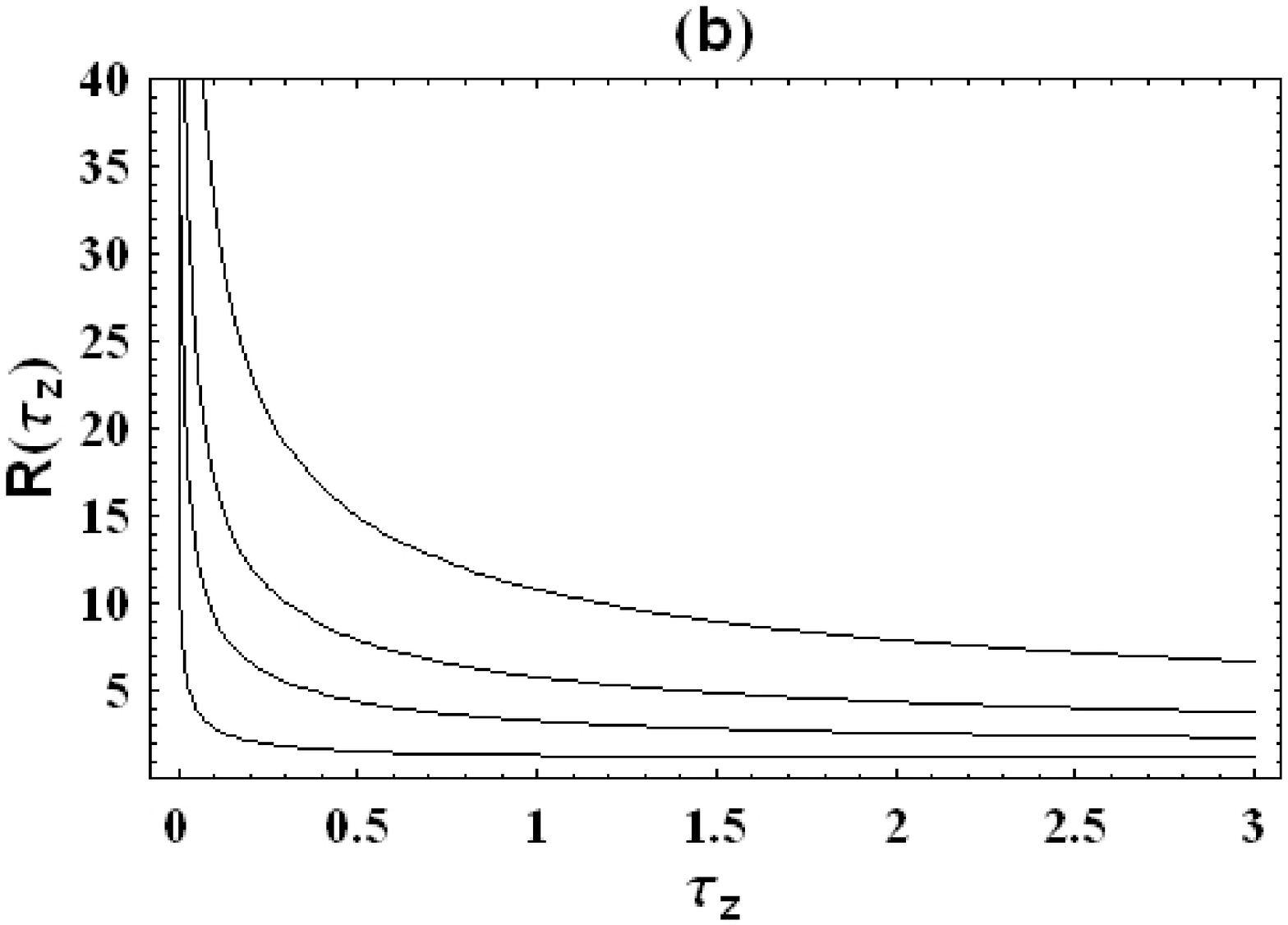}
\caption{(a) Relation between $R$ and the two control parameters
$(\tau_{z},\eta_{x})$.
 (b) Sectional views of (a), with $\eta_{x}=1,\ 5,\ 10,\ 20 $ from bottom to top.  The ratio $R$ is calculated
 from variable $\Delta q_{x}$ with $\Delta
k_{x}$ fixed at the origin.}
\end{figure}

We proceed to calculate the ratio for variable $\Delta q_{x}$, i.e.,
$R\equiv \delta \Delta q_{x}^{{\rm single}}/\delta \Delta
q_{x}^{{\rm coinc}}$, where the unconditional variance is obtained
from the single--particle observation as:
\begin{eqnarray}
\delta^{2} \Delta q_{x}^{{\rm single}}&=&\langle \Delta
q_{x}^{2}\rangle-\langle \Delta q_{x} \rangle^{2}\\
\nonumber &=&\int {\rm d}\Delta k_{x}{\rm d}\Delta q_{x} \Delta
q_{x}^{2}|A_{\frac{\pi}{2}}|^{2}\\ \nonumber &-&\left(\int {\rm
d}\Delta k_{x}{\rm d}\Delta q_{x}\Delta q_{x}|A_{\frac{\pi}{2}}|^{2}
\right)^{2},
\end{eqnarray}
and coincidence measurement gives the conditional variance at some
specified $\Delta k_{x}$:
\begin{eqnarray}
&& \delta^{2} \Delta q_{x}^{{\rm coinc}}=\langle \Delta
q_{x}^{2}\rangle_{\Delta k_{x}}-\langle \Delta q_{x} \rangle^{2}_{\Delta k_{x}}\\
\nonumber &&=\frac{\int {\rm d}\Delta q_{x} \Delta
q_{x}^{2}|A_{\frac{\pi}{2}}|^{2}}{\int{\rm d}\Delta
q_{x}|A_{\frac{\pi}{2}}|^{2}}-\left(\frac{\int {\rm d}\Delta
q_{x}\Delta q_{x}|A_{\frac{\pi}{2}}|^{2}}{\int{\rm d}\Delta
q_{x}|A_{\frac{\pi}{2}}|^{2}} \right)^{2}.
\end{eqnarray}
Substituting Eqs. (11)--(13) into the definition of $R$, we yield
$R(\eta_{x},\tau_{z})$ as a function of parameters $\eta_{x}$ and
$\tau_{z}$, the result of which is illustrated in Fig. 3 with
$\Delta k_{x}$ fixed at the origin. From that, one can see that the
entanglement increases monotonously when $\eta_{x}$ increases or
$\tau_{z}$ decreases, which indicates that higher entanglement can
be achieved by squeezing the linewidth of the incident photon or
broadening the wave packet of the atom. In particular, when
$\eta_{x}>1$, we have:
\begin{eqnarray}
R\approx
\frac{\eta_{x}+\sqrt{\frac{2}{\pi}}(1+\tau_{z})}{2\sqrt{\tau_{z}}},
\end{eqnarray}
from which it is found that the entanglement increases linearly with
$\eta_{x}$ and will be abruptly enhanced when $\tau_{z}$ tends to
zero. As a remark, we emphasize that all the conclusions above
 hold qualitatively the same either if $\Delta k_{x}$ is specified otherwise or one calculate the ratio $R$
 from the other variable $\Delta k_{x}$.\\

The ratio $R$, which can be obtained experimentally by comparing the
momentum dispersion variance, is an appropriate quantification for
the entanglement contained in the probability amplitude correlation
(thus can be seen as an evaluation of the ``amplitude
entanglement''). Next, we can see that it reveals a correct varying
tendency for the entanglement with its control parameters. However,
the definition of $R$ is dependent on its representation space and
different choices for the basis of Hilbert space will cause distinct
values of $R$. This is because we only use the amplitude of the
wavefunction to construct $R$, and then all entanglements
included in phase\cite{10 phase ent} is lost. \\

To obtain the ``total entanglement'', we calculate the Schmidt
number\cite{12 Schmidt num} and compare it with the entanglement
ratio $R$ in the following
section.\\

\section{Full entanglement in scattered photon }

Mathematically, for a bipartite system in pure state, the
entanglement of an unfactorable wavefunction can be completely
characterized by the Schmidt number, which is denoted by
$K\equiv(\sum_{n=0}^{\infty}\lambda_{n}^{2})^{-1}$, where
$\lambda_{n}'s$ are eigenvalues of the integral equation \cite{11
Schmidt dec}:
\begin{eqnarray}
\int {\rm d}\Delta k_{x}' \rho^{{\rm P}}(\Delta k_{x},\Delta
k_{x}')\phi_{n}(\Delta k_{x}')=\lambda_{n}\phi_{n}(\Delta k_{x}),
\end{eqnarray}
the density matrix for photon is defined as:
\begin{eqnarray}
\nonumber \rho^{{\rm P}}(\Delta k_{x},\Delta k_{x}')\equiv \int {\rm
d}\Delta q_{x} A_{\frac{\pi}{2}}(\Delta q_{x},\Delta
k_{x})A^{*}_{\frac{\pi}{2}}(\Delta q_{x},\Delta k'_{x}), \\
 \
\end{eqnarray}
where, note that we have taken away the time--dependent phase in the
density matrix since it does not contribute to entanglement.
Although we do it with the photon, Schmidt number can be equally
obtained through the atomic density matrix, and the eigenfunctions
of atom $\left[\psi_{n}(\Delta q_{x})\right]$ can be related to
those of photon through:
\begin{eqnarray}
\psi_{n}(\Delta q_{x})=\frac{1}{\sqrt{\lambda_{n}}}\int {\rm
d}\Delta k_{x}A_{\frac{\pi}{2}}(\Delta q_{x},\Delta
k_{x})\phi^{*}_{n}(\Delta k_{x}),
\end{eqnarray}
where $\phi_{n}(\Delta k_{x})$ and $\psi_{n}(\Delta q_{x})$
$(n=1,2\cdot\cdot\cdot)$ form complete orthonormal sets for the
photon and atom respectively. With these discrete modes, the
unfactorable wavefunction can be expanded into a sum of factored
products uniquely:
\begin{eqnarray}
A_{\frac{\pi}{2}(\Delta q_{x},\Delta
k_{x})}=\sum_{n}\sqrt{\lambda_{n}}\psi_{n}(\Delta
q_{x})\phi_{n}(\Delta k_{x}).
\end{eqnarray}
Then, the Schmidt number $K$, which is an estimation of the number
of modes that are ``important'' in making up the expansion of Eq.
(18), serves as a quantitive measurement of entanglement\cite{7
scattering}\cite{12 Schmidt num}. Note $K$ is independent from
representation since all $\lambda's$ keep the same in different
representations, thus can be seen as a quantity of the full
entanglement information (both amplitude
and phase entanglement) kept in the collective wavefunction.\\

\begin{figure}
\centering
\includegraphics[height=5cm]{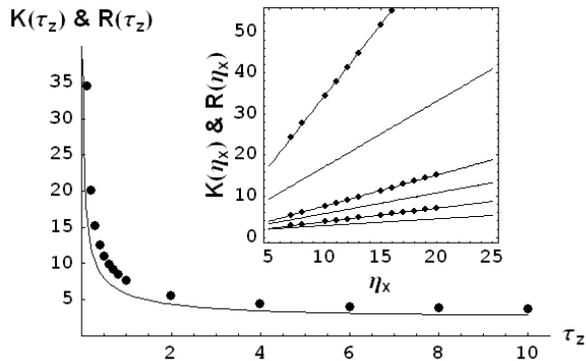}
\caption{Schmidt number $K$ and the amplitude entanglement degree
$R$ in dependence on $\tau_{z}$ with $\eta_{x}=10$. Spots are
numerical results for $K$ whereas solid line is plotted for $R$. The
inset shows them as functions of $\eta_{x}$ with $\tau_{z}$ fixed,
lines from bottom to top are depicted as: $R(\tau_{z}=10)$,
$K(\tau_{z}=10)$, $R(\tau_{z}=1)$, $K(\tau_{z}=1)$,
$R(\tau_{z}=0.1)$, $K(\tau_{z}=0.1)$, respectively. }
\end{figure}

Since Eq. (15) is not analytically solvable, we use a discrete
eigenvalue equation to approximate the integral equation. Up to a
reliable precision, we use $1000\times1000$ matrices to carry out
the diagonalization, and collect some of the results in Fig. 4,
where we also compare Schmidt number $K$
with the amplitude entanglement ratio $R$.\\

From the numerical results, we find that, similar to the ratio $R$,
$K$ rises linearly with parameter $\eta_{x}$ and will increase
rapidly when the linewidth of incident photon is squeezed narrower
to the atomic linewidth $\Gamma$, i.e, $\tau_{z}<1$; secondly, when
$\tau_{z}$ is fixed, the slope of $K(\eta_{x})$ is always larger
than
 that of $R(\eta_{x})$, which means that more entanglement information will transfer to phase when $\eta_{x}$ becomes larger,
 and this phenomena will become more evident when $\tau_{z}$
 is reduced, e.g., when $\tau_{z}=0.1$,
 $R\approx 1.58\eta_{x}+1.39$ whereas $K\approx 3.44\eta_{x}+0.08$, which indicates that more than half
 of the entanglement information will be unavailable by momentum dispersion observation when $\eta_{x}$
 goes large on this condition. \\

Another phenomena is notable, when $\tau_{z}=1$, i.e., the linewidth
of the incident photon is not squeezed and can be prepared directly
by spontaneous emission from the same atom, we find $K\approx
0.75\eta_{x}+0.16$ $(\eta_{x} \gg 1)$ whereas in the case of
spontaneous emission\cite{6 Singlephoton} $K\approx 0.28\eta +0.72$
$(\eta \gg 1)$. This difference indicates that, although in both
cases, entanglement is generated from momentum conservation in the
process of photon emission with atomic recoil, the absorption of the
incident photon will add some entanglement due to its coherent
pumping effect. As $K$ is linear with $\eta$ (or $\eta_{x}$), we
define the ``entanglement pumping coefficient'' as:
$${\rm EPC}\equiv \frac{{\rm slope\  of\ } K(\eta_{x}) {\rm\ in\
scattering}}{{\rm slope\ of\ } K(\eta ){\rm\ in\ spontaneous\
emission}}\ ,$$ since the constant term in $K(\eta)$ plays a minor
role when entanglement is large. The defined coefficient ${\rm EPC}$
shows the times that entanglement is increased by the coherent
pumping of an incident photon. As it is independent on the atomic
parameter, it reflects the ability of entanglement of the photon
separately. We collect some numerical results in Fig. 5 and fit it
with ${\rm EPC}\approx 1.1/\tau_{z}+1.5$ within $\tau_{z}\in(0,1)$,
from which, one sees that ${\rm EPC}$ increases rapidly when
$\tau_{z}$ diminishes, which also implies that, if the incident
photon is prepared monochromatically on its limit condition, i.e.,
$\tau_{z}\rightarrow 0$, the scattered photon will be highly
entangled to the recoiled atom.\\

We plot the amplitude of the first three Schmidt modes for the
photon with $\eta_{x}=10$ and $\tau_{z}=1$ in Fig. 6. We find that
their number of peaks in momentum space is proportional to the
Schmidt mode index, but the separations of different peaks are more
distinct than in the case of spontaneous
emission\cite{7 scattering}.\\

\section{transmitted photon}
\begin{figure}
\centering
\includegraphics[height=4cm]{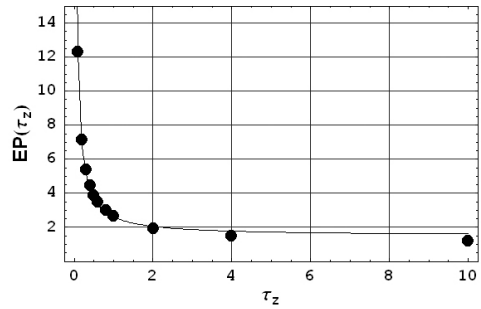}
\caption{Entanglement pumping coefficient ${\rm EPC}$ as a function
of $\tau_{z}$. The solid line is its fitted function
$1.1/\tau_{z}+1.5$.}
\end{figure}

To consider the transmitted photon, we make the observation angle
$\theta=0$, and yield the collective wavefunction from Eq. (10):
\begin{eqnarray}
A_{0}&=&-\chi_{0}G_{z}(q_{z})P_{z}(k_{z})\\
\nonumber &+&\chi_{0}\frac{\pi}{4}\left(\frac
{\Gamma}{ck_{0}}\right)^{2}
\frac{\tau_{x}\tau_{y}G_{z}(q_{z})P_{z}(k_{z})}{1-i(\Delta
k_{z}+\Delta q_{z}+\frac{\hbar k_{0}^{2}}{2m\Gamma})}.
\end{eqnarray}
One can see that, in Eq. (19), the first term describes that the two
particles are free of interaction and keep their initial factorable
wave form; the second term reflects the entanglement. Usually, the
second term is much smaller than the first one since $(\frac
{\Gamma}{ck_{0}})^{2}\ll1$, but one can enlarge it by choosing some
special physical system, such as the artificial atom with low
excited level and high coupling to its resonant modes. However, this
improvement can add few entanglement between the transmitted photon
and recoiled atom, because interference between the two terms in Eq.
(19) will weaken the correlation of the two particles at a great
deal. To make it clear, we show the contour and density plots for
the probability amplitude of $A_{0}$ in Fig. 2 on an artificial
condition $\frac{\pi}{4}(\frac{\Gamma}{ck_{0}})^{2}\tau_{x}\tau_{y}=
1$, and
yield $ R \approx  K<2$ in this situation.\\

The eigenfunctions of transmitted photon for the first three modes
with $\eta_{z}=10$ and $\tau_{z}=1$ are collected in Fig. 6, from
which one can see that, due to the interference, the corresponding
modes of the transmitted photon exhibit one peak less than that of
the
scattered photon.\\
\begin{figure}
\centering
\includegraphics[height=7cm]{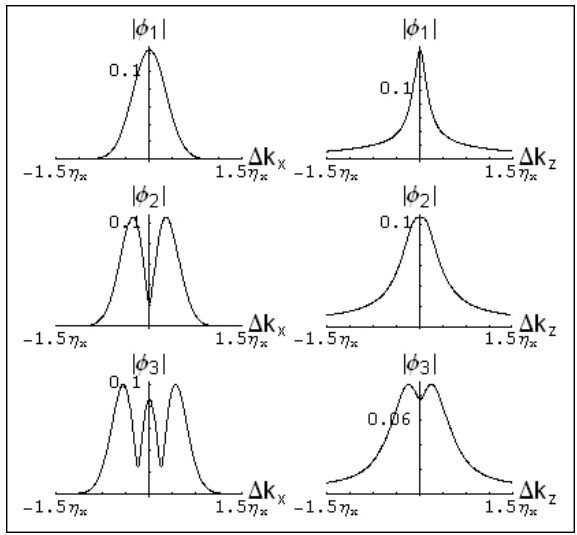}
\caption{First three Schmidt modes for the scattered and transmitted
photon. Left column is for the scattered photon with $\tau_{z}=1$
and $\eta_{x}=10$; right column is for the transmitted photon with
$\tau_{z}=1$, $\eta_{x}=10$, and
$\frac{\pi}{4}(\frac{\Gamma}{ck_{0}})^{2}\tau_{x}\tau_{y}= 1$ for
illustration.}
\end{figure}

\section{conclusion}
We analyze the physically fundamental interaction between a single
photon and a free artificial atom in vacuum. With a few physical
approximations, the general solution of the photon--atom wave
function is obtained, from which, it is found that the initially
uncorrelated particles will evolve to be entangled due to momentum
conservation in scattering. To evaluate the entanglement in the
scattering, firstly, we use an experimentally accessible parameter
$R$, which denotes the ratio between momentum variance in
single--particle and in coincidence observations, and yield its
simple dependences on the two physical control parameters
$\eta_{x}\equiv \frac{\delta q_{x}\hbar k_{0}}{m\Gamma} $ and $
\tau_{z}\equiv \frac{\delta k_{z}}{\Gamma/c}$; secondly, we use
standard Schmidt decomposition to reveal the full entanglement
information and find out its varying tendency similar to that of
$R$, which indicates that high entanglement can be achieved by
either squeezing the linewidth of the incident photon or broadening
the scale of atomic wave packet. Furthermore, compared with
spontaneous emission, we defined a parameter ${\rm EPC}$ to evaluate
the entanglement enhancement due to the coherent pumping effect of
the resonant incident photon. In the end, we found out that, for the
transmitted photon, one
can expect little entanglement due to the interference between the transparent and scattered wave.\\

\section*{ACKNOWLEDGMENTS}
One of the authors (HG) acknowledges J. H. Eberly for his
discussions when drafting this manuscript. This work is supported by
the National Natural Science Foundation of China (Grant No.
10474004), and DAAD exchange program: D/05/06972 Projektbezogener
Personenaustausch mit
China (Germany/China Joint Research Program).\\


\begin{references}
\bibitem{statement}  See, e.g., Ref[5], the artificial atom is implemented which can provide high
 coupling coefficient and so can potentially be applied in our proposed experimental scheme.
\end{references}
\end{document}